\documentstyle[aps,epsf,prd,floats,cite,preprint]{revtex}

\preprint{\vbox{ \hbox{MIT-CTP-3229} \hbox{hep-ph/0201056} }}
\title{Localized Fermions and Anomaly Inflow \\
 via Deconstruction }
\author{Witold Skiba and David Smith}
\address{
      Center for Theoretical Physics, Massachusetts Institute of Technology,
      Cambridge, MA 02139 \vspace{.3cm}}
\begin{document}

%Create the title page
\maketitle
\begin{abstract}

We study fermion localization in gauge theory space. We
consider four dimensional product gauge groups in which light chiral
fermions transform under different gauge factors of the product group.
This construction provides a suppression of higher dimensional operators.
For example, it can be used to suppress dangerous proton decay operators.
The anomalies associated with the light chiral fermions are compensated 
by Wess-Zumino terms, which in the continuum limit reproduce the five
dimensional Chern-Simons term. 

\end{abstract}
%%%%%%%%%%%%%%%%%%%%%%%%%%%%%%%%%%%%%%%%%%%%%%%%%%%%%%%%%%%%%%%%%
\tighten
\newpage
%%%%%%%%%%%%%%%%%%%%%%%%%%%%%%%%%%%%%%%%%%%%%%%%%%%%%%%%%%%%%%%%%
%Main body of the paper
\section{Introduction}
%%%%%%%%%%%%%%%%%%%%%%%%%%%%%%%%%%%%%%%%%%%%%%%%%%%%%%%%%%%%%%%%%
Extra-dimensional scenarios provide a novel mechanism of generating
small coefficients~\cite{NimaMartin}. The overlap of spatial
wave-functions can be small for fields localized at different positions.
Operators involving those fields can thus be suppressed. In an effective
four-dimensional description such small coefficients appear to be accidental.
There are several important applications of this idea, for example
explaining proton stability in theories with a low fundamental scale,
or modeling small Yukawa couplings of the Standard Model fermions.

In this article we study fermion localization in four dimensions.
Localization in extra dimensions is replaced by localization in ``theory
space'' using the recent notion of deconstruction~\cite{ACG1,HPW}.
Deconstruction trades extra dimensions for lattices of four dimensional
gauge theories. The sites of such lattices are gauge groups, so
extra dimensions are replaced by products of gauge groups.
The product gauge group is spontaneously broken to a single gauge group, which
is the lowest Kaluza-Klein (KK) mode of the extra dimensional gauge group.
The massive gauge bosons corresponding to the broken generators play the role
of the massive KK tower.  

Deconstruction is interesting for several reasons. First, it provides a
UV completion of higher dimensional theories. Second, it motivates interesting
model building. It is certainly not surprising that a space-time dimension
can be successfully discretized. When the lattice spacing is small, lattice
theories should provide a sufficiently accurate description of their
continuum counterparts. What is interesting for model building  is that 
some extra-dimensional features are preserved by very coarse lattices.
When the number of lattice sites is small the spectrum does not resemble
higher dimensional theories at all. Several models motivated
by deconstruction illustrate this point. Examples include novel extensions
of the Standard Model~\cite{ACG2,CHW}, mechanisms for communicating
supersymmetry breaking~\cite{CEGK,CKSS}, low scale
unification~\cite{accelerated}, and models for breaking the unified
gauge symmetries~\cite{orbifoldGUTS}. In some cases, there exist
higher dimensional theories which guide four-dimensional
model building, in other cases four dimensional models cannot be extended
into higher dimensions.

Deconstruction has yielded some formal developments as well. These
include a description of little string theories in terms of gauge 
theory~\cite{ACKKM}, a single gauge group description of extra dimensions in
the large number of colors limit~\cite{IraWitek}, and models of noncommutative
geometry~\cite{Alishahiha,noncom}. Other works investigated warped
background geometry~\cite{Sfetsos}, topological objects in field
theory~\cite{Hill}, Seiberg-Witten curves~\cite{instantons},
and even ventured into gravity~\cite{Bander}. One of our goals is to
investigate anomalies associated with chiral fermions localized in theory
space and to derive the Chern-Simons term in the continuum limit.

The article is organized as follows. In the next section we describe
how to localize fermions in a product group theory. Our mechanism
relies on decoupling massive fermions and it is not a lattice description 
of fermionic zero modes trapped on topological defects.
In Section~\ref{sec:fermions}, we also explain how small coefficients
are generated and why high-energy scattering is 
suppressed~\cite{NimaYuvalMartin} in our framework.
In Section~\ref{sec:anomaly},
we consider the continuum limit of a theory with localized fields.
Since chiral fermions transform under different gauge groups, the theory
appears to be anomalous. The anomalies are canceled by the Wess-Zumino (WZ)
terms. In the continuum limit, the WZ terms associated with different
lattice sites combine to reproduce the bulk Chern-Simons term. 
In Section~\ref{sec:models}, we present supersymmetric models that realize
our idea, paying particular attention to proton decay operators. The models
are extensions of the minimal supersymmetric Standard Model (MSSM), in which
light quarks and leptons transform under different gauge groups.

%%%%%%%%%%%%%%%%%%%%%%%%%%%%%%%%%%%%%%%%%%%%%%%%%%%%%%%%%%%%%%%%%
\section{Localized Fermions}
\label{sec:fermions}
%%%%%%%%%%%%%%%%%%%%%%%%%%%%%%%%%%%%%%%%%%%%%%%%%%%%%%%%%%%%%%%%%

In this section we introduce the construction of localized
fermions, which is central to the rest of the article. We then explore
a few consequences of such localization. We will not summarize the
inner workings of deconstruction and the emergence of extra dimensions 
since it has been described extensively in the literature and instead
we refer the reader to Refs.~\cite{ACG1,HPW}.

A useful notation for deconstructed theories is the moose
notation~\cite{moose}, in which individual gauge groups are denoted
as circles. Fermionic (scalar) fields are denoted by solid (dashed)
oriented lines. A line incoming to a circle indicates a field transforming
in the fundamental representation, outgoing one in the antifundamental.
Consequently, lines connecting two circles represent fields transforming
under two gauge groups in the ``bifundamental'' representation, that is
fundamental under one of the gauge groups, antifundamental under
the other.

%%%%%%%%%%%%%%%%%%%%%%%%%%
\begin{figure}[!hb]
\epsfxsize=9cm
\hfil\epsfbox{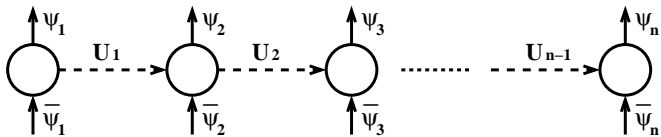}\hfill
\caption{Moose diagram for localized fermions. The circles represent
$SU(k)$ gauge groups, solid lines stand for Weyl fermions in the fundamental
and antifundamental of $SU(k)$, while the dashed lines represent scalars.}
\label{fig:one}
\end{figure}
%%%%%%%%%%%%%%%%%%%%%%%%%%

Consider an $SU(k)^n$ product gauge group with a pair of Weyl
fermions, $\psi_i$ and $\bar{\psi}_i$, for each group, as illustrated in
Fig.~\ref{fig:one}. In addition to the fermion fields, there are
bifundamental scalar ``link'' fields, $U_i$. The link fields obtain vevs,
$v$, proportional to the identity, which break $SU(k)^n$ to the diagonal
subgroup $SU(k)$. We do not define the dynamics that provides the
vevs for the link fields. It could either be arranged by a choice of
potential for the link fields or emerge dynamically if the link fields
are condensates of more fundamental objects.

In addition to gauge interactions, we add the following mass terms and
Yukawa couplings.
\begin{equation}
\label{eq:couplings}
  {\cal L}=\lambda \sum_{i=i}^{n-1} \bar{\psi}_i U_i \psi_{i+1} +
           \mu \sum_{i=2}^{n} \bar{\psi}_i \psi_{i},
\end{equation}
where $\lambda$ is a common Yukawa coupling and $\mu$ a common mass term.
Note that we have omitted a mass term for $\bar{\psi}_1$ and $\psi_1$.
When the link fields obtain vevs, the Yukawa interactions become 
mass terms with mass $M=\lambda v$. In what follows it will be crucial
that $\mu < M$. Our objective for now is to explain how localization works and
we postpone to Section~\ref{sec:models} the obvious questions on how to obtain
such a pattern of couplings, which includes the hierarchy of masses,
and the absence of Yukawa couplings $\bar{\psi}_{i+1} U^\dagger_i \psi_i$.  

From Eq.~(\ref{eq:couplings}) the mass matrix for fermions is
\begin{equation}
\label{eq:masses}
  {\cal M} = \left(\begin{array}{cccc} \bar{\psi}_1 & \bar{\psi}_2 &
                      \ldots & \bar{\psi}_n \end{array} \right)
             \left(\begin{array}{cccccc} 0 & M & 0 & \ldots & 0 & 0 \\
                                         0 & \mu & M & \ldots & 0 & 0 \\
                   \vdots & \vdots & \vdots & \ldots & \vdots & \vdots \\
                   0 & 0 & 0 & \ldots & \mu & M \\
                   0 & 0 & 0 & \ldots & 0 & \mu \end{array} \right)
             \left(\begin{array}{c} \psi_1 \\ \psi_2 \\
                      \vdots \\ \psi_n \end{array} \right).    
\end{equation}
It is apparent from the construction that $\psi_1$ is massless
irrespectively of the values of $\mu$ and $M$ since it does not couple
to any other fermion. There must also be another massless fermion,
$\bar{\psi}$, because at most $n-1$ $\psi$'s are massive. When $\mu=0$, 
$\bar{\psi}_n$ is massless, while all other fermions have mass $M$.
So in this simple case massless fermions transform under the first
and the last $SU(k)$ factors of the moose.

When $\mu \neq 0$, the zero modes are $\psi_1$ and 
\begin{equation}
      \bar{\Psi}= \frac{ (\frac{\mu}{M})^{n-1} \bar{\psi}_1 -
                           (\frac{\mu}{M})^{n-2} \bar{\psi}_2 + \ldots
                           + (-1)^{n-1} \bar{\psi}_n }{
               \sqrt{1+(\frac{\mu}{M})^2 +\ldots + (\frac{\mu}{M})^{2(n-1)}}}.
\end{equation}
We assume that $(\frac{\mu}{M})^{n-1} \ll 1$, so that $\bar{\psi}_1$
appears with a small coefficient in the linear combination comprising
the $\bar{\Psi}$ zero mode. Thus, operators involving both zero modes will
generally be suppressed as their overlap is small. The coupling suppression
is the result of a symmetry present for special values of parameters.
When $\mu=0$, there is a $U(1)$ global symmetry that rotates $\bar{\psi}_n$
only and does not act on any other field. For $\mu \neq 0 $ this symmetry
is approximate.

When unknown or poorly understood dynamics becomes relevant at a high
energy scale, $\Lambda$, one needs to regard the theory as an effective theory
with a cutoff $\Lambda$. At $\Lambda$, all operators allowed by
gauge symmetries can be generated, presumably with order one coefficients.
We are interested in such operators if they involve the zero modes 
since these are the operators that are accessible to experiment.
High energy dynamics can generate, for example, the following
gauge invariant combinations in our moose theory
\begin{equation}
 \bar{\psi_1} \psi_1, \; \; 
 \frac{1}{\Lambda^{n-1}} 
 \bar{\psi_n} U^\dagger_{n-1} \ldots U^\dagger_1 \psi_1.
\end{equation}
Expressing these operators in terms of the zero modes and replacing the
link fields by their vevs reveals the suppressions associated with them.
The first operator is small due to the $\bar{\Psi}$ zero mode
wave-function, the second one is suppressed if $v<\Lambda$. Therefore,
these gauge invariants always bring in small coefficients
\begin{eqnarray}
  \bar{\psi_1} \psi_1 & \rightarrow & (\frac{\mu}{M})^{n-1} \bar{\Psi} \psi_1, 
             \nonumber \\
  \frac{1}{\Lambda^{n-1}} 
   \bar{\psi_n} U^\dagger_{n-1} \ldots U^\dagger_1 \psi_1 
                      & \rightarrow & (\frac{v}{\Lambda})^{n-1}
                                      \bar{\Psi} \psi_1.
\end{eqnarray}
We will apply these results to models of quark-lepton separation in
Section~\ref{sec:models}.

Another interesting consequence of localized fermions is the suppression
of high-energy scattering cross-sections~\cite{NimaYuvalMartin}. The $t$
and $u$ channel scattering amplitudes turn out to be subdued compared to
the low-energy results when their natural scaling with $1/t$ or $1/u$ is
factored out. In extra dimensions the suppression appears for momentum
transfers comparable to the inverse of the separation distance and becomes
an exponential suppression in the large momentum limit. The moose
theory resembles a five dimensional theory only at distances shorter
than the compactification scale, but longer than the lattice spacing.
At distances shorter than the lattice spacing, the theory exhibits four
dimensional behavior. Therefore, in the deconstructed theory the suppression
will reach its maximum near the energies corresponding to the inverse lattice
spacing and will not increase after that.

To illustrate this point we study the scattering 
\begin{equation}
  \psi_1 + \bar{\Psi} \longrightarrow \psi_1 + \bar{\Psi}
\end{equation}
in the tree-level approximation. When the link fields get vevs the
spectrum of the gauge bosons includes
a massless multiplet of the diagonal $SU(k)$ and $n-1$ massive
adjoints of $SU(k)$. Let us assume that all the gauge couplings are
equal to $g$. Then, in terms of the original gauge bosons, $A_i$, the mass
eigenstates $a_j$ are~\cite{HPW}: 
\begin{eqnarray}
  a_0 & = & \sqrt{\frac{1}{n}} (A_1+A_2+\ldots + A_n), \nonumber \\ 
  a_j & = & \sqrt{\frac{2}{n}} \sum_{i=1}^n 
                               \cos\left(\frac{(2 i-1) j \pi}{2 n}\right) A_i, 
\label{eq:gaugebosons}
\end{eqnarray}
where $j=1,\ldots, n-1$. The masses of the vector bosons are
$m^2_j= 2 g^2 v^2 \sin^2(\frac{j \pi}{2 n})$. We express the gauge
couplings of $\psi_1$ and $\bar{\Psi}$ in terms of the mass eigenstates.
The mass eigenstates defined in Eq.~(\ref{eq:gaugebosons}) can be summarized
as 
\begin{equation}
  A_i = \sum_{j=0}^{n-1} \gamma_{ij} a_j, \, 
                    \gamma_{ij}=\sqrt{\frac{2}{n (1+\delta_{j,0})}}
                       \cos\left(\frac{(2 i-1) j \pi}{2 n}\right).
\end{equation}
The amplitude for the scattering can be computed by summing up the
exchanges of all mass eigenstates
\begin{eqnarray}
  {\cal A}(\psi_1 + \bar{\Psi} \rightarrow \psi_1 + \bar{\Psi}) 
      & \propto & \sum_{j=0}^{n-1} \sum_{i=1}^n 
       \frac{(\mu/M)^{2(n-i)}}{1+(\mu/M)^2+\ldots+(\mu/M)^{2 (n-1)}}
         \gamma_{ij} \gamma_{1j} \frac{1}{t-m_j^2} \nonumber \\
      &\rightarrow & (\frac{\mu}{M})^{2 (n-1)} \frac{1}{t}, \; \; 
                     {\rm when} \; \; t \gg m_j^2,
\end{eqnarray}
where the last identity follows from the orthogonality of $\gamma_{ij}$.
We have assumed that $\mu \ll M$ and kept only the lowest order term
in $\frac{\mu}{M}$.

This result for the suppression of the amplitude could have been anticipated.
At high energy it is more convenient to work with the original gauge
bosons, $A_i$, rather than the mass eigenstates $a_j$. Since $\psi_1$ couples
only to $A_1$, the sum over all mass eigenstates can be replaced by the
exchange of $A_1$ at high energies, and the amplitude suppression arises
simply from suppression of the coupling of $A_1$ to $\bar{\Psi}$.
The amplitude for the inclusive scattering 
$\psi_1 + \bar{\Psi} \longrightarrow \psi_1 + X$ is down by only a factor of
$(\mu/M)^{n-1}$ since at sufficiently high energies massive fermions
can be produced. The massive $\bar{\psi}$'s couple more strongly to $A_1$ than
$\bar{\Psi}$ couples to $A_1$, so the suppression comes only from the initial
state. If the fermions are heavier than the gauge bosons ($\lambda > g$)
there could be an interesting intermediate regime, in which the inclusive
cross-section is highly suppressed and it is less suppressed at both lower
and higher energies.

%%%%%%%%%%%%%%%%%%%%%%%%%%%%%%%%%%%%%%%%%%%%%%%%%%%%%%%%%%%%%%%%%
\section{Continuum limit and anomaly inflow}
\label{sec:anomaly}
%%%%%%%%%%%%%%%%%%%%%%%%%%%%%%%%%%%%%%%%%%%%%%%%%%%%%%%%%%%%%%%%%
In this section we study the issue of anomaly cancellation in
an effective theory obtained by decoupling heavy fermions. 
For simplicity, we consider the non-supersymmetric theory described
in the previous section, where we set $\mu=0$ in
Eq.~(\ref{eq:couplings}).  We choose the Yukawa couplings to be larger
than the gauge couplings, so we can study the theory after the fermions
are integrated out, while all gauge bosons are still present in the
effective theory.

Fermions that obtain masses via the Yukawa couplings leave well-known
non-decoupling remnants in the effective theory. The classic examples
are the Goldstone-Wilczek current~\cite{GoldstoneWilczek} and the
WZ term~\cite{WZ}. The link fields in Fig.~\ref{fig:one} transform
as $(\bar{k},k)$ under the neighboring gauge groups. When the link fields
obtain vevs the setup is identical to that considered in
Refs.~\cite{WZ,DHokerFarhi}. In the moose theory we obtain a gauged WZ
term for each link field.

It is important that we consider the appropriate energy regime for this
derivation. At the highest energy, the theory is non-anomalous since
there is a vector-like pair of fermions for each gauge group. After
integrating out fermions we need to include WZ terms involving the 
link fields and the gauge bosons. If we integrated out all massive gauge
bosons there would be no WZ term left. The massless theory, described by
the diagonal subgroup of $SU(k)^n$, is not anomalous. It contains two
massless Weyl fermions with opposite gauge charges. 

The gauged parts of the WZ terms~\cite{Aneesh,A-GG} summed over all link
fields are
\begin{eqnarray}
\label{eq:WZ}
  {\cal L}_{\rm eff}= & \frac{1}{48 \pi^2} \sum_{i=1}^{n-1} & 
       \left\{  {\rm tr}\left[(A_i dA_i + dA_i A_i)(U_i A_{i+1} U^\dagger_i +
               U_i dU^\dagger_i) +
                (dU_i U^\dagger_i dA_i U_i A_{i+1} U_i^\dagger) 
               -{\rm p.c.}\right] \right\} \nonumber \\
      & &+\left\{ {\rm tr}\left[A_i^3(U_i A_{i+1} U^\dagger_i + 
               U_i dU^\dagger_i)
               -(A_{i+1} U_i^\dagger dU_i A_{i+1} U_i^\dagger A_i U_i)
               -{\rm p.c.}\right] \right\} \nonumber \\
      & &- \left\{ \frac{1}{2} {\rm tr}
               \left[A_{i+1} U_i^\dagger A_i U_i A_{i+1} U_i^\dagger A_i U_i
               \right] \right\} \\
      & &+\left\{ {\rm tr}\left[A_i(dU_i U_i^\dagger)^3
               +\frac{1}{2} A_i dU_i U_i^\dagger A_i dU_i U_i^\dagger
               +U_i A_{i+1}U_i^\dagger dU_i U_i^\dagger dU_i U_i^\dagger
               -{\rm p.c.} \right] \right\}. \nonumber
\end{eqnarray}
In the above equation p.c.\ denotes the parity conjugate under which
$A_i \leftrightarrow A_{i+1}$ and $U_i \leftrightarrow U_i^\dagger$.
The parity conjugate applies to all the terms inside each square bracket.
Meanwhile, $d$ denotes the four dimensional derivative and all terms
are contracted with the four dimensional $\epsilon$ tensor even though
we omitted Lorentz indices.

After integrating out $\psi_i$ and $\bar{\psi}_{i+1}$ the Lagrangian is not
gauge invariant, and the addition of the $i$-th WZ term from Eq.~(\ref{eq:WZ})
compensates for that. Under gauge transformations $\omega_i$, $\omega_{i+1}$ 
the variation of the $i$-th WZ term is
\begin{equation}
 \delta {\cal L}_{\rm eff}^i= \frac{-1}{24 \pi^2} {\rm tr}
   \left[\omega_i (dA_i dA_i +\frac{1}{2} d(A_i^3))-(i \rightarrow i+1)\right].
\end{equation}
It is clear that the gauge variations of the neighboring WZ terms cancel
for their common gauge group. That is, the sum of $i$ and $i+1$ WZ terms
is invariant under the gauge transformation $\omega_{i+1}$.
In the full effective Lagrangian the cancellation
occurs for all gauge groups, except for the endpoints. This is, of course,
what the sum of the WZ terms is supposed to do in the effective theory. 
Only at the endpoints are there light chiral fermions, which are not
gauge invariant on their own.

We now consider the continuum limit, in which the lattice spacing
$a \rightarrow 0$. We evaluate only terms of order $a$ and neglect
terms ${\cal O}(a^2)$. We can then identify $A(y)=A\equiv A_i$, where
$y$ is the fifth coordinate. Therefore, $A_{i+1}=A+a \partial_5 A$ and
$U_i=1+a A_5$. The lowest order terms are linear in $a$, so we can pull
$a$ out and convert the sum over all lattice sites into an integral
$a \sum_i \rightarrow \int dy$.

The terms in Eq.~(\ref{eq:WZ}) are grouped such that the terms inside
the first curly bracket contribute only to $A dA dA$ and $A^3 dA$,
inside the second one to $A^3 dA$ and $A^5$, while inside the third one
only to $A^5$. The terms inside the last curly bracket are all
${\cal O}(a^2)$ and do not contribute.

From now on $A$ denotes the five dimensional (5D) vector boson,
while $d$ denotes 5D derivatives. Terms with five vector indices are all
contracted with the 5D $\epsilon$ tensor. After a bit of algebra we obtain 
\begin{eqnarray}
\label{eq:CS}
  {\cal L}_{\rm eff}&=&  \frac{-1}{24 \pi^2} \int d^4 x dy 
                      \,\, {\rm tr}\left[AdAdA + \frac{3}{2} A^3dA + 
                      \frac{3}{5} A^5 \right] \\
                    && +\int d^4 x dy \, ({\rm total\ 4D\ derivatives}). \nonumber
\end{eqnarray}
The first line in the above equation is precisely the 5D Chern-Simons
term~\cite{A-GG}. The total derivative terms act on either $A^4$ or
$A^2 dA$. After integrating by parts these terms do not contribute since at
large distances they decay at least as fast as $1/r^4$. 

The Chern-Simons term is what compensates for the anomaly with localized
fermions in the 5D theory~\cite{CallanHarvey,Naculich}. The effect of
gauge transformations on fermions localized at the endpoints of the 5D
space is canceled by the anomaly inflow from the bulk of five
dimensions. It is an interesting check of the deconstruction procedure
that the continuum limit of the moose theory reproduces properly the
Chern-Simons term. If the moose theory did not reproduce the Chern-Simons
term it would be very puzzling.

%%%%%%%%%%%%%%%%%%%%%%%%%%%%%%%%%%%%%%%%%%%%%%%%%%%%%%%%%%%%%%%%%
\section{Models}
\label{sec:models}
%%%%%%%%%%%%%%%%%%%%%%%%%%%%%%%%%%%%%%%%%%%%%%%%%%%%%%%%%%%%%%%%%

Here we use the localization mechanism introduced in Section II to
construct extensions of the MSSM that feature effective quark-lepton
separation. In these models dangerous dimension-five proton decay operators
are suppressed by gauge invariance.

How is proton stability maintained in the ordinary MSSM? Unlike the situation
in the Standard Model, baryon number is not an accidental symmetry of the
renormalizable interactions. For example, $Q D L$ is an allowed operator in
the superpotential. Such unwanted interactions are usually prohibited by
imposing R-parity. Although a global symmetry, R-parity is likely to be
gauged in the underlying theory; in the models considered here R-parity will
arise as an unbroken discrete subgroup of $U(1)_{B-L}$ gauge symmetry.

However, it is well known that R-parity alone is not sufficient for stabilizing
the proton, because there exist R-parity conserving non-renormalizable
operators that violate baryon and lepton number. For example, proton decay
bounds require the effective suppression scale $\Lambda$ in  the superpotential
operator, ${1\over \Lambda} QQQL$ to be at least $\sim 10^{26}$ GeV, whereas
naively, without invoking small couplings, one would expect this operator to
have at most $\sim M_{Planck}^{-1}$ suppression.  Similarly, the operator
$UUDE$ must be suppressed by at least $\sim 10^{22}$ GeV.  Thus, we see a
motivation for considering effective quark-lepton separation in supersymmetric
theories even if the cutoff is quite high.  

There were three crucial ingredients required in the localization framework
described in Section~\ref{sec:fermions}.
First, there were no mass terms for fermions transforming under the
first gauge group. Second, Yukawa couplings acted only in
one direction, for example coupling fermions from left to right but
not the other way around. Third, there was a hierarchy between
$\mu$ and $M=\lambda v$ such that $(\frac{\mu}{M})^{n-1}$ was small.
We will address these issues in turn.

Explicit mass terms are absent in any chiral theory. Assuming that
one of the endpoints of the moose diagram has the field content of the
Standard Model chirality takes care of the first requirement.

The second requirement is trivially satisfied in models with two lattice
sites. In models with more lattice sites, supersymmetry can address this
requirement. Because the superpotential terms are holomorphic,
hermitian conjugates of the link fields cannot appear in the
superpotential. 

This brings us to the hierarchy of masses.  In supersymmetric theories,
the choice of a small value for the parameter $\mu$ is guaranteed to be
radiatively stable. In the models we construct, $\mu$ will appear in front
of operators like ${\overline L}L$ and ${\overline E}E$
and will be on similar footing to the $\mu$ parameter appearing in front
of $H_U H_D$ in the superpotential.

%%%%%%%%%%%%%%%%%%%%%%%%%%
\begin{figure}[!ht]
\epsfxsize=5cm
\hfil\epsfbox{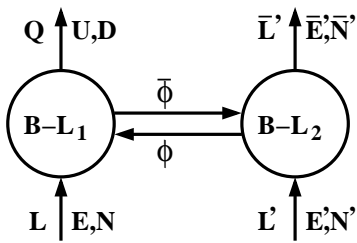}\hfill
\caption{Moose diagram for the $MSSM \times U(1)_{(B-L)}\times U(1)_{(B-L)'}$
model.}
\label{fig:two}
\end{figure}
%%%%%%%%%%%%%%%%%%%%%%%%%%

Our simplest model contains just two lattice sites. Models with additional
lattice sites allow the successful gauge coupling unification of the MSSM
to be preserved, or the cutoff scale to be lowered far beneath the 
Planck scale.

The model with two lattice sites is depicted in Fig.~\ref{fig:two}. The gauge
group is taken to be the product of the  Standard Model gauge group with two
additional $U(1)$'s, which we call $U(1)_{(B-L)}$ and  $U(1)_{(B-L)'}$.
The idea is that the light quarks and leptons will be separated in
``$B-L$ space,'' rather than in a space having to do with the Standard Model
gauge group. Three generations of MSSM matter fields transform under
$U(1)_{(B-L)}$ alone with the usual $B-L$ charges.  Thus we have 
$Q(1/3,0)$, $U(-1/3,0)$, $D(-1/3,0)$, $L(-1,0)$, $E(1,0)$, and $N(1,0)$,
where we list charges under $U(1)_{(B-L)}\times U(1)_{(B-L)'}$ 
in parenthesis.  Meanwhile, transforming under $U(1)_{(B-L)'}$ alone are
three generations of vector-like pairs $L'(0,-1)+\overline{L'}(0,1)$,
$E'(0,1)+\overline{E'}(0,-1)$, and  $N'(0,1)+\overline{N'}(0,-1)$, again
with ordinary $B-L$ charge assignments, and with the usual transformation
properties under the Standard Model gauge group. The Higgs doublet superfields
$H_U$ and $H_D$ are neutral under $U(1)_{(B-L)} \times U(1)_{(B-L)'}$, so
they are spread out over $B-L$ space, and Yukawa couplings may be written
down at both lattice sites.

We assume that $U(1)_{(B-L)} \times U(1)_{(B-L)'}$ is broken to its
diagonal subgroup by vevs of link fields $\phi(1,-1)$ and
${\overline \phi}(-1,1)$, which are singlets under the Standard Model
gauge group. This breaking can be arranged through the superpotential term
$S(\phi {\overline \phi}-v^2)$, where $S$ is some singlet chiral superfield. 
The Yukawa interactions 
\begin{equation}
   \lambda_L L \phi \overline{L'} 
     +\lambda_E E \overline{\phi} \, \overline{E'} 
     +\lambda_N N \overline{\phi} \, \overline{N'}
\end{equation}  
and the $\mu$ terms
\begin{equation}
  {\mu}_L L'  \overline{L'} +{\mu}_E E'  \overline{E'}+{\mu}_N N'  \overline{N'}
\end{equation} 
give the leading contributions to the masses for the lepton superfields. In
the absence of electroweak symmetry breaking, there is a  massless zero mode
\begin{equation}
   l={L'-{\mu_L \over M_L} L \over \sqrt{1+({\mu_L \over M_L})^2}},
\end{equation} 
where $M_L=\lambda_L v$.
Similarly, the zero modes $e$  and $n$ are linear combinations of $E$ and $E'$,
and $N$ and $N'$, respectively.  

To have effective quark lepton separation, there are two requirements we
must impose.  The first is that $v \ll \Lambda$, where $\Lambda$ is the cutoff
scale suppressing non-renormalizable operators. This ensures, for instance,
that the operator
\begin{equation} 
  {1 \over \Lambda}Q Q Q {\overline{\phi}\over \Lambda} L'
\label{eq:direct}
\end{equation}
yields a $QQQL'$ operator with strong extra suppression $v/\Lambda$. 
The second requirement is $\mu \ll M$ for $L$ and $E$.  This ensures that
the zero modes $l$ and $e$ will essentially be equal to $L'$ and $E'$.  
Then, for example, the operator
\begin{equation}
  \frac{1}{\Lambda} Q Q Q L
\end{equation}
contains a $Q Q Q l$ coupling suppressed by the additional small
factor $\mu_L/M_L$.

Taking $\Lambda \sim M_{Planck}$, we require $v/\Lambda$ and $\mu/M$
both to be less than roughly $10^{-8}$ to evade proton decay bounds.  
For example, one could have $v\sim 10^{11}$ GeV and set the $\mu$'s
to be $\sim$ TeV, roughly equal to Higgs $\mu$ parameter of the MSSM.

As things stand, the model still has an unbroken $U(1)_{B-L}$, which can be
broken relying on standard methods.  For instance, we can introduce $X$ and
$\overline{X}$ superfields, with charges (0,2) and (0,-2) under
$U(1)_{(B-L)} \times U(1)_{(B-L)'}$, and neutral under the Standard Model
gauge group, and add the superpotential couplings
\begin{equation}
    T(X \overline{X} - v_{B-L}^2)   + X N' N',
\end{equation} 
where $T$ is a singlet chiral superfield.  The vevs of $X$ and $\overline{X}$
then break the diagonal $U(1)_{B-L}$ to a discrete subgroup that contains
R-parity. Taking $v_{B-L} \sim 10^{15}$ GeV  gives roughly the right size
seesaw neutrino masses to account for atmospheric neutrino oscillations.

How low can the cutoff be taken in this model?  Taking $\mu_L$ and  $\mu_E$
to zero leaves only the contributions to $QQQl$ and $UUDe$  coming from the
coupling through link fields, as in Eq.~(\ref{eq:direct}). 
Requiring $v>$TeV then leads to the bound $\Lambda>10^{14}$ GeV\@.  
The cutoff scale can be reduced further below the Planck scale in models
with additional sites.  In fact, there is a completely different motivation
for considering models with larger numbers of sites, coming from gauge
coupling unification.  In the model with two sites, the scale $v$ is
forced to be below  $\sim 10^{11}$ GeV, leaving us with extra matter at
this scale that spoils the successful gauge coupling unification in the MSSM.
By taking more sites, the upper bound on $v$ coming from proton decay
is raised, and with enough sites $v$ can be raised to the GUT scale.

%%%%%%%%%%%%%%%%%%%%%%%%%%
\begin{figure}[!ht]
\epsfxsize=8cm
\hfil\epsfbox{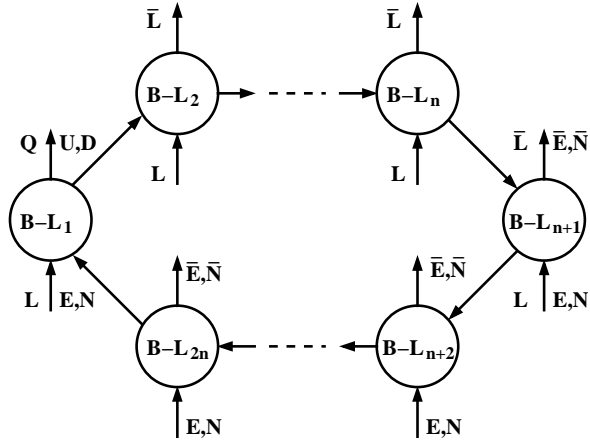}\hfill
\caption{Moose diagram for the $MSSM \times U(1)_{(B-L)}^{2n}$
model.}
\label{fig:three}
\end{figure}
%%%%%%%%%%%%%%%%%%%%%%%%%% 

Consider the following extension of the two-site model, shown in
Fig.~\ref{fig:three}:  the gauge group is the product of the Standard Model
group with $2n$ additional $U(1)$ factors, called $U(1)_{(B-L)}^i$ for $i$
running from 1 to $2N$.  Link fields $\phi_i$ have charges $(1,-1)$
under $U(1)_{(B-L)}^i\times U(1)_{(B-L)}^{i+1}$ (with the index $2n+1$
identified with 1), and acquire vevs that break the product of $U(1)$'s to
their diagonal subgroup. To cancel mixed
$U(1)_{(B-L)}^i\times U(1)_{(B-L)}^{i+1}$ anomalies~\footnote{We thank 
Erich Poppitz for bringing these to our attention.} we introduce 
$\overline{\phi}_i(-1,1)$ fields that do not acquire vevs. Alternatively,
one could have eight copies of link fields $\phi_i(1,-1)$
and one copy of $\bar{\Phi}_i(-2,2)$. Quark-lepton separation is 
preserved even if all link fields acquire vevs since Yukawa couplings
with $\bar{\Phi}_i$ are forbidden.
At site 1 three generations of MSSM matter
superfields reside as before.  For sites $2$ through $n+1$ we have
vector-like pairs $L+\overline{L}$ at each site,  while for sites
$n+1$ through $2n$ we have both $E +\overline{E}$ and 
$N + \overline{N}$ pairs at each site.   Suppose for simplicity that
we have universal $\mu$ terms for $L$, $N$, and $E$
at each site where they appear, and universal Yukawa couplings to the link
fields, $\lambda$.  Then, as before, we require  $\mu \ll M=\lambda v$ and
$v \ll \Lambda$.  The zero modes $l$, $n$, and $e$ are then dominantly
$L_{n+1}$, $N_{n+1}$, and  $E_{n+1}$, respectively.  
The higher-dimensional operator
\begin{equation}
  {1 \over \Lambda^{N}}Q_1 Q_1 Q_1 \phi_1 \phi_2 ...\phi_n L_{n+1}
\end{equation}   
leads to a $QQQl$ coupling suppressed by an extra factor $(v/\Lambda)^n$. 
For example, taking $n=3$ and $\Lambda=M_{Planck}$ allows us to take $v$ near
the GUT scale.  Alternatively, for fixed, small $v$, $\Lambda$ can be lowered
by taking $n$ large. Meanwhile, the contribution to $QQQl$ coming from
\begin{equation}
  {1\over \Lambda}Q_1 Q_1 Q_1 L_1
\end{equation} 
is suppressed by an extra factor $(\mu/M)^n$ (identical suppressions
arise for $UUDe$). Thus, even a mild hierarchy between $\mu$ and $M$ can
lead to a sufficient suppression of proton decay if one makes $n$ large enough.

%%%%%%%%%%%%%%%%%%%%%%%%%%%%%%%%%%%%%%%%%%%%%%%%%%%%%%%%%%%%%%%%%
\section{Conclusions}
%%%%%%%%%%%%%%%%%%%%%%%%%%%%%%%%%%%%%%%%%%%%%%%%%%%%%%%%%%%%%%%%%
In theories with extra dimensions fermion localization can be used to 
suppress certain operators. One of the most striking results
obtained using this mechanism is that the cutoff scale
can be taken quite low without violating proton decay bounds,
by spatially separating quark fields and  lepton fields.
The anomalies associated with the separated chiral fermions are canceled
in this framework by the anomaly inflow associated with the five-dimensional
Chern-Simons term.

In this paper we have deconstructed this picture by building
four-dimensional models with fermions localized in gauge theory space. 
The localization results from particular patterns of couplings and is not
due to topological features. Our low-energy theory contains light chiral
fermions transforming under different factors of a product gauge group.
The anomaly associated with these fermions is canceled by the Wess-Zumino
terms generated by integrating out heavy fermions. We checked that in the
continuum limit these Wess-Zumino terms combine to form the Chern-Simons term.

We used deconstruction to obtain four-dimensional models with
effective quark-lepton separation.  In the models we constructed,
the product gauge group is comprised of factors of $U(1)_{(B-L)}$.
The degree of separation (and thus the level of suppression of dangerous
proton decay operators) depends on the number of lattice sites, and
on the ratios of $v/\Lambda$ and $\mu/(\lambda v)$. Here,
$v$ is the scale of spontaneous breaking to the diagonal
$U(1)_{(B-L)}$, $\Lambda$ is the UV cutoff, $\mu$ is the mass between
vector-like pairs at a given site, and $\lambda$ is the Yukawa coupling
involving the link fields and fermions at adjacent sites. 
Models with very few sites require large hierarchies among these mass scales; 
models with only mild hierarchies require taking a larger number
of sites.

One model building challenge is localizing the matter fields without 
localizing the Higgs doublets, so that both $Q D H_D$ and $L E H_D$ are
present. This is accomplished trivially in our models because the Higgs fields 
are neutral under the ``moose'' $B-L$ gauge groups. Neutral fields are
naturally spread out over the whole theory space since they can couple at any
site without suppression.  In this setup, the subject of flavor is essentially
set aside.  One could construct alternative models in which different
generations of quarks and leptons have different ``profiles'' in the gauge
space (see extra-dimensional models of Ref.~\cite{fermionmasses}). 
Moreover, one could localize the Higgs fields, e.g.\ in theories where the
Standard Model gauge group is discretized, so that certain Yukawa couplings
are suppressed by factors of $v/\Lambda$ required to link the Higgs
to the relevant fermions.  In this way, a connection could be drawn
between suppressed proton decay and small Yukawa couplings in this framework.

\acknowledgements
       
We would like to thank Nima Arkani-Hamed, Andy Cohen, Lisa Randall,
Ira Rothstein, and Martin Schmaltz for discussions. This work was
supported by the Department of Energy under grant DE-FC02-94ER40818.

%%%%%%%%%%%%%%%%%%%%%%%%%%%%%%%%%%%%%%%%%%%%%%%%%%%%%%%%%%%%%%%%%%%%%

\end{document}